\documentstyle[aps,pre,epsfig,multicol]{revtex}
\begin{document}
\draft
\title{Non-local conservation in the coupling field: effect on critical dynamics}
\author{Parongama Sen}

\address{Department of Physics, Surendranath College, 24/2 Mahatma Gandhi Road,
Calcutta 700009, India}

\maketitle
\begin{abstract}
We consider the critical dynamics of a system with  an  $n$-component
non-conserved order parameter  coupled to a conserved field with long
range diffusion. An exponent $\sigma$ characterizes the long range
transport, $\sigma=2$ being the known locally conserved case.
With   renormalisation group calculations done upto one loop order, 
several regions are found with
different values of the dynamic exponent $z$  in the $\sigma -n$ plane. 
For $n<4$, there are three regimes, I: nonuniversal, $\sigma$
dependent $z$, II: universal with $z$ depending on $n$ and III':
conservation law irrelevant, $z$ being equal to that in the
nonconserved case. The known locally conserved case belongs to regions
I and II.

\end{abstract}
\bigskip

\begin{multicols}{2}
  Conservation laws play significant roles in critical dynamics in
  thermodynamic systems.  The dynamics of such systems have been
  studied over the last several decades \cite{bray,hh,gunton}.  These
  conservations, as constraints, generally distinguish the various
  dynamic universality classes for a given equilibrium class.  For
  example, in Ising ferromagnets the order parameter may not be
  conserved (model A of ref. \cite{hh}) though it is so for a binary
  mixture (model B).  In antiferromagnets or binary alloys undergoing
  an order-disorder transformation, a secondary parameter, the total
  magnetization, is conserved (model C) \cite{hh,eisen}.  These
  classes are characterized by the dynamic exponent $z$ that connects
  the divergence of the time $(\tau)$ and length $(\xi)$ scales near
  the critical point, $\tau \sim \xi^z$.  For several cases
  \cite{bray,hh,hhm2} renormalisation group arguments give exact
  result for the dynamic exponent $z$.  However, the numerical studies
  give somewhat different values which do not always agree with each
  other\cite{sds,ds}.

Most of the earlier studies assume local conservation of the conserved
field, though local conservation is not a necessary, but a sufficient,
condition for  conservation in general.  Recently the issue of
non-local conservation has been addressed in the case of conserved order
parameter (model B) either by relaxing local diffusion or allowing
infinite range transport\cite{bray1,TK,satya,moseley,AB,ruten}.  
For the nonconserved case, numerical simulations have shown that the
dynamic exponent can depend on the nature of the dynamics also
\cite{salas}.  It is therefore important to determine the robustness
of the wellknown dynamic universality classes.

Allowing long range diffusion in the Ising model means that the
Kawasaki spin exchange now takes place with further neighbour spins
also.  
The case where the range of the exchange becomes infinite corresponds to what
we shall call "global" conservation.
For the conserved order parameter case, model B, the effect of
global conservation seems to be drastic. In fact the dynamic
correlation functions of the $q$ state Potts model has been shown to
be analogous to that the Ising model with fixed magnetisation and long
range interactions (i.e., essentially model B with non-local conservation) 
\cite{satya}.  Dynamics for local and global
conservation are also expected to be different \cite{ruten}.  For
$O(n)$ symmetric order parameter, the conservation law has been found
to be irrelevant for global conserved case (i.e.  infinite range
Kawasaki dynamics) \cite{bray,bray1} and the dynamic exponent $z$ is
argued to be the same as that of model A.  The situation is expected
to be more subtle when a conserved quantity is coupled to a
nonconserved order parameter as for example in model C.

In the case of model C, the dynamics of the nonconserved order
parameter is slowed down compared to model A because of its coupling
to a scalar nonordering conserved field. Even though the conserved
field is not critical, its coupling to the order parameter induces
long range correlations.  These singular correlations in turn affect
the relaxational dynamics, yielding $z_A < z_C < z_B$ where the
subscript denotes the model name.  With local conservation, model C
has several regions in the $d-n$ plane (where $d$ is the spatial
dimension and $n$ the number of components of the order parameter).
There are  three different regions corresponding to different values of
the exponent $z$.  Above $ n = 4$, the behaviour is model A-like with
$z = 2 + c\eta$, i.e, there is correction to $O(\epsilon^2)$,\footnote{The equilibrium exponents $\alpha$, $\nu$ and $\eta$
are obtained from the scaling of the specific heat ($\sim |r|^{-\alpha}$),
correlation length ($\sim |r|^{-\nu}$), and correlation function
($\sim  |{\bf{x}}|^{-(d - 2+\eta )}$ at $r = 0$) respectively, where $r$ is the deviation
from a critical point.} 
where $\epsilon=4-d$ . 
For $ 2<n < 4$, the values of $z$ are $2 + 2\alpha/(n\nu)$, and $z=2 +
\alpha/\nu$ for $ n < 2$, where $\alpha$ and $\nu$ are respectively
the specific heat and the correlation length exponents. (These values
of $z$ are for $\alpha > 0$, otherwise one 
takes $\alpha = 0$ and $z$ once again becomes $z_A$.)  The early time
behaviour of    dynamical models  with short range initial correlations also gives
a new exponent \cite{zheng,OJ,cardy,sutapa}.  Now, the emergence of the long
range correlation in the conserved field, the hallmark of model C, is
an equilibrium property.  Then, are the universal dynamic behaviours,
as reflected in the $d-n$ plane, sensitive to the nature of the
dynamics? We study this problem by considering a long range transport
for the conserved field, a convenient way of interpolation between the
local to the global conservation.

Let us consider the dynamics of an $n$ component order parameter field
$\phi$ coupled to a scalar noncritical but conserved field $m$ which
may be the energy density or the magnetisation in the Ising
antiferromagnetic case or particle density in the binary alloy case,
or annealed mobile impurities.  We allow for subdiffusive or
superdiffusive transport of the conserved field.  This means long
range transport for particles or the impurities \cite{cond9809175} or
long range exchange of spins in magnetic systems.  The Hamiltonian for
model C reads,
\begin{eqnarray}
H = \int d^d{\bf x}[\frac{1} {2} r\phi^2({\bf x}) + \frac{1}{2}(\nabla
\phi ({\bf x}))^2 
+\tilde u\phi^4({\bf x}) \nonumber \\
     + \gamma \phi^2({\bf x})m ({\bf x})+\frac{1}{2}C^{-1}  
m^2 ({\bf x})],
\end{eqnarray}
where, $\phi^2({\bf x}) = \sum_{i=1}^{n}\phi_i^2({\bf x}), \phi^4({\bf x}) =
 [\sum_{i=1}^{n}\phi_i^{2}({\bf x})]^2, r=0$ is the mean field critical point, and $C>0$ ensures
noncriticality of $m$.

For purely relaxational dynamics, the equation of motion obeyed by
$\phi$ is 
\begin{mathletters}
\begin{equation}
 \frac {1}{\Gamma}\frac {\partial {\phi_\alpha ({\bf x},t)}}{\partial t} =  
- \frac {\partial {H}}{\partial \phi_\alpha ({\bf x},t)}
+ \eta_\alpha ({\bf x},t),
\end{equation}
while $m$ satisfies the following dynamical equation in Fourier space
\begin{equation}
\label{eq:2}
 P(k) \frac {\partial {m ({\bf k},t)}}{\partial t} =
   -\frac { \partial {H}}{\partial m (-{\bf k}, t)}
+ \zeta ({\bf k},t)
\end{equation}
\end{mathletters}
with the noise obeying 
\begin{mathletters}
\begin{eqnarray}
\langle \zeta  \rangle = 0~,&& ~~~ \langle \eta_\alpha  \rangle = 0,\\
\langle \eta_\alpha  ({\bf x},t)\eta_{\alpha '} ({\bf x}',t') \rangle &=& 2\Gamma
\delta({\bf x}-{\bf x}')\delta(t-t')\delta_{\alpha,\alpha '}\\
\langle \zeta  ({\bf k},t)\zeta({\bf k}',t') \rangle &=&  2
P(k)\delta({\bf k}+{\bf k}')\delta(t-t').
\end{eqnarray}
For  model C with  local conservation  
one chooses, $P(k) = 1/(\lambda k^2)$. Here we use the   general form 
\cite{bray1}   
\begin{equation}
\label{eq:1}
P(k) = 
 [\frac {1}{\lambda_\sigma  k^\sigma }  + \frac {1}{\lambda_0}]
\end{equation}
\end{mathletters}
 
For conservation to hold one strictly needs to satisfy the continuity
equation only. For local conservation, the additional assumption is
that the current is determined by the {\it local} change in the
chemical potential (this is the case when $\sigma = 2$ and Fick's law
is obeyed, i.e, $j({\bf x}) \propto \nabla \mu({\bf x})$, where $j$ is
the current density and $\mu({\bf x})$ is the chemical potential in the
particle picture).  
When the conservation is non-local, Fick's law is no longer valid.
This is analogous to the invalidity of Ohm's law for diffusing electrons 
following Levy flight instead of nearest neighbour hopping \cite{Arkhi-No}.
In the long range diffusion case, instead of
Fick's law, one can have a more general constitutive equation $j({\bf
  x}) \propto \int \kappa ({\bf x}-{\bf x}')\nabla \mu({\bf x}')d{\bf x}'$
(which gives back Fick's law for $\kappa = \delta({\bf x}-{\bf x}')$).
This for a power law decay of $\kappa ({\bf x})$ on Fourier
transformation gives Eq.  (\ref{eq:2}) with Eq. (\ref{eq:1}).  We
further include the momentum independent term, a kinetic coefficient,
$\lambda_0$ which is generated by the RG itself.  A value of $\sigma
\ne 2$ simulates the non-locality. The only constraint on $\sigma $ to
ensure conservation 
 is $\sigma > 0$. The limit $\sigma \rightarrow 0^+$ is the case corresponding
to global conservation.

We use  the momentum shell renormalisation group method
where the Fourier components $\phi_k(t)$ for the hard  modes $\Lambda
/b < k < \Lambda$  
are eliminated 
and the  scale changes $k = k/b$ and $t = b^zt$ are made. We
eventually set $\Lambda=1$. The long distance long time results are
obtained from the flow of the renormalized parameters.
The static fixed points (at one loop) of this model are easily obtained
$(\epsilon=4-d)$ 
\cite{hhm2} as follows:
 \begin{mathletters}
\begin{eqnarray}
 r^* &=& \frac {1} {2} \frac {n+2} {n+8} \epsilon  \\
 u^* &=& \frac {\epsilon } {4K_d (n+8)}
\end{eqnarray}
(here  $ u = \tilde u -\frac {1}{2} \gamma^2C$, and $K_d =
2^{1-d}\pi^{-d/2}\Gamma (d/2)$).    
\begin{equation}
\label{eq:3}
 (\gamma^2 C)^* = \frac {\tilde \alpha } {2n\nu K_d}
\end{equation}
And either $ \gamma^* =  0$ 
or
\begin{equation}
 \tilde \alpha /\nu = (4-n)\epsilon/(n+8); ~~~~~\gamma ^* \ne 0  
\end{equation}
\end{mathletters}
$\tilde \alpha $ is the exponent of the temperature dependence of the
largest term in the specific heat, i.e, $\tilde \alpha = \alpha$ if
$\alpha > 0$ and $\tilde \alpha = 0$ otherwise ($\alpha$ is the
specific heat exponent).  For $ n > 4$, $\gamma ^* = 0$ is the only
fixed point. In general, $\gamma^* = 0$ is the fixed point for $n >
n_c$ where $n_c$ also has an $\epsilon$ expansion.

Considering the various vertex functions, the renormalisation equation
for  the transport and kinetic coefficients can be obtained \cite{hhm2}. 
The RG
transformation being analytic cannot renormalise the singular terms of
$P(k)$, as in the local case. However the transport coefficient
generates (and renormalizes) the kinetic coefficient, $\lambda_0$, and
also renormalizes the kinetic coefficient of the order parameter.
Defining the dimensionless parameter $\mu_{\sigma}=\Gamma
C/\lambda_{\sigma}$, the recursion relations at the static fixed point
of Eq.~(\ref{eq:3}) can be written as (choosing thin shells $b = 1
+\delta l$ with $\delta l\rightarrow 0$)
\begin{mathletters}
\begin{eqnarray}
\label{eq:4}
\frac {\partial \Gamma^{-1}}{\partial l} &=& \Gamma^{-1}[-z+2 + \frac
{2\tilde \alpha/\nu n}{(1+\mu^{-1})}],  \\
\label{eq:6}
\frac {\partial \lambda_0^{-1} } {\partial l}& =&   \lambda_0^{-1} 
(-z+\tilde \alpha/\nu + 
   \frac
{2n\gamma^2\lambda_0 K_d  }{\Gamma}) \\  
\label {musigma}
\frac{\partial \mu_\sigma }{\partial l} &=& \mu_\sigma [\sigma  -2  +
\tilde \alpha/\nu - \frac{2\tilde \alpha/\nu n}{1+\mu^{-1}}].  
\end{eqnarray}
\end{mathletters}
Here $\mu = \mu_\sigma + \mu_0$, with 
 $\mu_0 = \frac {\Gamma C}{\lambda_0}$.
The above one loop contributions in the dynamic quantities  come  from the
$\gamma$ vertex.
(It maybe noted that we have excluded   $\eta $
 as $\eta = 0$ to first order in $\epsilon$. This is not crucial as
there is a cancellation in model C in higher order terms\cite{hhm2,hh}.)
 
The dynamic exponent $z$  is chosen to keep $\Gamma $ scale
invariant.  Eq.~(\ref{eq:4}) then gives
\begin{equation}
  \label{eq:5}
  z= 2 + \frac{2\tilde \alpha/\nu n}{(1+\mu^{-1})}.
\end{equation}
The fixed point of $\mu_0$ is obtained from eqs. (\ref{eq:6}) and
(\ref{eq:3})
$$\mu_{0}^* = \frac {\tilde \alpha/\nu} {2(z-\tilde \alpha/\nu)}$$

The fixed points of $\mu_\sigma$ are therefore $0, \infty$ or a non-zero
finite value given by  
\begin{equation}
\label {mstar}
\mu_{0}^* +\mu_{\sigma}^* = \frac {\sigma -2 +\tilde \alpha/\nu}{2-\sigma +\frac 
{\tilde \alpha}{\nu} \frac {2-n}{n}}
\end{equation} 

\medskip

{\it Stability  of fixed points and exponents: }

Let us analyse the stability of the fixed points  
for the different values of $\sigma$. 

Above $ n = 4$, $\gamma^* =0$ is the only fixed point and therefore
the conserved field gets decoupled in long lengthscale limit.  This is
nothing but the model A fixed point where the ordering and
non-ordering fields are no longer coupled and therefore $z = z_A$.  To
recover $z_A= 2+ c \eta$ one needs to go to higher loops, where the $u$
vertex starts contributing.

$\mu_\sigma = 0$ is the fixed point valid for $\sigma < 2 -\alpha/\nu$
and the exponent $z = 2 + O(\epsilon^2)$ for all $n$ .  At this fixed
point the conservation of the coupling field simply becomes
irrelevant.  One therefore again recovers model A exponent $z=z_A$,
although $\gamma ^* \ne 0$.

We note that as $\mu_{\sigma}^*$ cannot be negative  and $\tilde
\alpha  \sim O(\epsilon )$, 
  (\ref{mstar})  will give a physical   fixed point only
if $ n < 2/(1+p)$ where  $ p = (\sigma - 2) / (\tilde \alpha/\nu)$
and $p > -1$. 
For $4 > n > 2/(1+p), ~ p > -1$,  
$\mu_\sigma = \infty$ will be the stable   fixed point.
The latter gives $ n = 4$ at $p = -0.5$, so for $ -1 < p <-0.5$,
we find that  a non-zero finite fixed point for  $\mu_{\sigma}^* $ 
is always stable. 
In particular, for  $\sigma = 2$,   we get back the
results of \cite{hh}: $ z = 2 + \tilde \alpha /\nu $ 
for $n < 2$, $z = 2 + 2\tilde \alpha /n\nu $ for $2 < n < 4$ and $z =
2 + O(\epsilon^2)$ 
for $n > 4$.

\medskip

Thus we have the following  regions in the $\sigma -n$ plane:\\\\
I. $~~~z = \sigma  +
\tilde \alpha/\nu$.\\ 
For\\
  $\sigma = 2 + p\tilde \alpha/\nu,  n < 2/(1+p);~ p > -1$ \\\\
II. ~$z = 2 + 2\tilde \alpha/n\nu$.\\
 For \\
 $\sigma =   2 + p \tilde \alpha /\nu,   4>  n > 2/(1+p);~ p > -1/2$\\\\
III.  $z = 2 + O(\epsilon^2)$, $\gamma^* = 0$.\\
For all $\sigma, n > 4$.\\\\ 
III'  $ z = 2 + O(\epsilon^2)$, $\gamma^* \ne  0$.\\
For $\sigma  < 2 - \alpha /\nu ~~~({\rm i.e.,}~ p = -1), n < 4.$\\

These are shown schematically in Fig. 1.  
We may add that there are discontinuities 
of $z$ only at the boundaries $\sigma = 2-\tilde \alpha /\nu$ and 
$n = 4$.  
The model C like  regions for $\sigma=2$ are in regions I and
II with nothing special for the $\sigma = 2$ line. The value of $z$ 
is also continuous across the boundary of regions I and II. 
 Of the two regions,
I is nonuniversal with a $\sigma$ dependent 
 dynamic exponent  $z = \sigma  + \tilde
\alpha/\nu$ while II is robust and independent of $\sigma$.
In  region II, $z$ is dependent on $n$ explicitly
while in region I, $n$ dependence is through the 
equilibrium exponents only.  
As one crosses the $\sigma = 2$ line,   region II
extends towards smaller values of $n$ for $\sigma > 2$ and
the opposite happens for $\sigma < 2$ where it vanishes at 
$n = 4, \sigma = 2 -\tilde \alpha/2\nu$ and region I survives
as long as $\sigma > 2 -\tilde \alpha /\nu$.  
  
The general $\sigma $ case, in addition, allows for a new regime where
the conservation law becomes 
irrelevant even though the coupling survives $(\gamma \neq 0)$ in the
long length scale 
limit and induces long range correlations in equilibrium.  The global
conservation case belongs to this class.
\medskip

Since the
nonrenormalizability of $\lambda_{\sigma}$ is valid to all orders, we
expect, as in the local model C, the dynamic exponent $z$, as obtained 
here, for regions I and
II to be {\it exact}. 
The above analysis is best for $d $ close to 4 as we have
done calculations upto $O(\epsilon )$ only. 
For $d $ close to 4, in
fact, region I exists for values of $\sigma \ge 2 $ and $\sigma < 2$
and very close to 2 only as $\alpha \sim O(\epsilon )$.  Continuing
with this result in lower dimensions, one expects region I to exist
over a larger range of $\sigma (> 2)$ so that the line separating
regions I and II in the positive half of the $\sigma - 2$ axis in Fig.
1 goes higher up for lower dimensions.  However, the line separating
regions II and III (or III' and III) in Fig 1 is shifted towards lower
values of $n$ for lower dimensions as $\alpha $ becomes negative at a
critical value of $n = n_c$ where $n_c$ also has an $\epsilon$ expansion.
(Fig. 1 has been drawn for $d$ close to  4).
In the $d-n $ plane therefore, region I is in general
depleted for a value of $\sigma$ higher than 2.
This depletion is maximum for 
$d = 4$.  
For $\sigma < 2 - \tilde \alpha/\nu$, only region III or III' survive
with $z = z_A$ for all $d$.    

Some of our results are in contrast with the long range
diffusion in model B, where $z$ is explicitly dependent on $\sigma$ as
long as the conservation is relevant.  The dynamical exponent is in
general expected to increase for a system with constraint, and
therefore $z \ge z_A$.  However, global conservation (i.e., $\sigma \rightarrow
0^+$) becomes irrelevant and one recovers $z = z_A$ in this limit
\cite{bray1,AB,moseley}.  This would the case for $\sigma < \sigma_c =
z_A -2 +\eta=O(\epsilon^2)$ for model B.
In model C, we find region III', where conservation is irrelevant, to be 
significantly larger with $\sigma_c=2 -\tilde \alpha/\nu$.  Even when the
conservation is relevant, we find a strongly universal regime, region
II, where the conservation slows down the dynamics but the nature of the
dynamics is irrelevant ($\sigma$ independence).

It is possible 
to implement this long range transport in a
simulation. The infinite range would correspond to $\sigma \rightarrow
0^+$, while any other $\sigma$ can be implemented by choosing the two
spins to be flipped, in the Kawasaki dynamics, at a distance $l$ with
probability $P(l) \sim l^{-(r+\sigma)}$ for large $l$.  However, it is
fairly easy to identify the irrelevance of the conservation in model B
numerically when the model A exponent is recovered, as $z_A$ and $z_B$
are easily distinguishable even with moderate level of precision in
calculations.  For model C, very high precision simulations are required
to verify our RG based scenario as  the
model C exponents in region I and II are  different from that of  model A
 by a much lesser margin.

 To summarise our results, we find the effect of long range diffusion
in model C with an $n$-component order parameter  to be highly significant. 
Characterising the long range
diffusion by an exponent $\sigma > 0$, several regimes with different
dynamical exponents $z$ are obtained in the $\sigma-n$ plane. 
For $n < 4$, a  region  with a  $\sigma$ dependent $z$  is found to exist 
as well as a region with strong universality where 
$z$ is independent of $\sigma$.
Most remarkable is the existence of a  region for $\sigma $ greater than
a critical value  
where the coupling with the field survives in the large length scale limit
but the conservation becomes irrelevant and $z= z_A$.
Such a region  is also present in model B with non-locally
conserved order parameter,  although for a much smaller range of
$\sigma$.      
For $n > 4$, the coupling with the field is irrelevant
and $z = z_A$ as expected.
The known results for the locally conserved model C are recovered with 
$\sigma = 2$.

\narrowtext
\begin{figure}
\psfig {file = 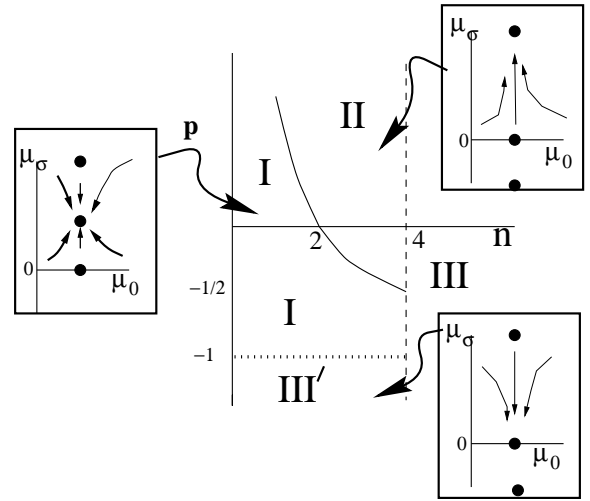, width=3in}
\vskip 0.5in
\caption{Regions corresponding to different values of $z$ are
shown for $p = \frac {\sigma - 2}{\tilde \alpha/\nu}$ and $n$. 
The flow diagrams in $\mu_\sigma -\mu_0$ plane are also shown. 
The flow is towards a finite non-zero value of $\mu_\sigma$ in region I,   
towards $\mu_\sigma \rightarrow \infty$ in region II and $\mu_\sigma = 0$
in region III'. }  
\end{figure}
\end{multicols}
\end{document}